# The martingale Z-test

Kenneth D. Harris, UCL Queen Square Institute of Neurology, London WC1N 3BG, UK

kenneth.harris@ucl.ac.uk, July 4, 2022.

We describe a statistical test for association of two autocorrelated time series, one of which generated randomly at each time point from a known but possibly history-dependent distribution. The null hypothesis is that at each time point, the two variables are independent, conditional on history until that time point. We define a test statistic that is a martingale under the null hypothesis and describe an asymptotic test for it based on the martingale central limit theorem. If we reject this null hypothesis, we may infer an immediate causal effect of the randomized variable on the measured variable.

In many fields of science, we would like to draw inferences from timeseries data over which we have partial control. In neuroscience, for example, an experimenter might present a subject with sensory stimuli and rewards over a series of trials, while observing their behavioral responses and brain activity. The experimenter controls the stimuli and rewards, and can deliver them randomly, with probabilities that may depend on the subject's history of choices. The subject's behavior and brain activity will depend not just on current conditions, but also on the history of stimuli and rewards, as well as unobservable internal variables that are serially correlated across trials. Classical statistical methods that require independent, identically distributed samples, cannot be used in such a situation[1].

This note describes a statistical test that can test a causal effect of one experimental variable, which is randomly generated on each trial from a possibly history-dependent distribution, on another measured variable, whose distribution is unknown and may be serially correlated. The test is based on the theory of martingales: random timeseries in which the assumption of statistical independence is relaxed, but for which a version of the central limit theorem still holds. By defining a timeseries which is a martingale under the null hypothesis, we can compute a test statistic which has a standard Gaussian distribution under the null, and so test the null hypothesis using a Z-test.

## History-conditional independence

Before describing the test, we distinguish two different notions of statistical independence for time series. The strong definition of independence between two time series $X_t$ and $Y_t$, which we term complete independence, means that $X$ and $Y$ are independent as vectors:

$$\mathbb{P}[X_1, \ldots, X_T, Y_1, \ldots, Y_T] = \mathbb{P}[X_1, \ldots, X_T]\mathbb{P}[Y_1, \ldots, Y_T]$$

The individual time series may be autocorrelated: $X_t$ may correlate with $X_u$ for any pair of times $t$ and $u$, but $X_t$ must be independent of $Y_u$.

The weaker definition is independence conditional on history. Define $\mathcal{H}_t$ to be the history up to time $t$, meaning the full set of observed values $X_1, \ldots, X_t, Y_1, \ldots, Y_t$, together with any additional unobserved variables during this time interval (the mathematical term for such a set of histories is a *filtration*). We say that $X$ and $Y$ are independent conditional on history if for each time $t$, the values of $X_t$ and $Y_t$ are independent given the history before then:

$$\mathbb{P}[X_t, Y_t | \mathcal{H}_{t-1}] = \mathbb{P}[X_t | \mathcal{H}_{t-1}]\mathbb{P}[Y_t | \mathcal{H}_{t-1}]$$

It is possible for a pair of timeseries to be independent conditional on history but not completely independent. We consider a simple example, related to the behavioral task used by the International Brain Laboratory[2]. On each trial of this task, a sensory stimulus is presented on the left or right, which we encode as a variable $A_t$ taking the value $\pm 1$. The probability of the two possible stimuli are not equal, and depend on time: in some trials, $A_t = +1$ with 80% probability, in other trials $A_t = -1$ with 80% probability. The probability switches at random times between blocks in which left or right stimuli are more probable. The subject must choose left or right, and is rewarded for a choice $C_t = \pm 1$ that matches the stimulus; a subject who can see the stimulus can thus get a reward on every trial. But a subject who cannot see the stimulus can still do quite well: by recalling which actions were rewarded on recent trials, they can infer the current block and thus choose the side is most likely to be rewarded. If a subject takes this strategy, then the timeseries of actions and stimuli will not be completely independent, since the action at time $t$ depends on the history of stimuli, actions, and rewards up to time $t-1$. However, if the subject cannot see the stimulus then the choice $C_t$ is independent of the stimulus $A_t$ conditional on the history $\mathcal{H}_{t-1}$, since the subject can only glean information about $A_t$ once they have observed whether the action $C_t$ was rewarded.

The martingale Z-test tests the null hypothesis that two timeseries are independent conditional on history.

## What is a martingale?

A martingale, crudely speaking, is a random numerical timeseries which does not show predictable drift. More precisely, a martingale is a timeseries whose expected value at time $t+1$, conditional on the history $\mathcal{H}_t$ up to time $t$, is the same as its actual value at time $t$:

$$\mathbb{E}[S_{t+1} | \mathcal{H}_t] = S_t$$

Our first example of a martingale is an unbiased random walk. If $X_1, X_2, \ldots$ are a series of independent random variables with mean 0, then their cumulative sum $S_t = \sum_{u=1}^{t} X_u$ is a martingale. The $X_t$ do not need to be identically distributed: if $\mathbb{E}[X_{t+1}|\mathcal{H}_t] = 0$ for all $t$, then $\mathbb{E}[S_{t+1}|\mathcal{H}_t] = S_t$, regardless of the distribution of $X_{t+1}$.

In fact, the $X_t$ do not even need to be statistically independent provided $\mathbb{E}[X_{t+1}|\mathcal{H}_t] = 0$. A timeseries $X_t$ obeying this condition is known as a martingale difference sequence, and its cumulative sum is a martingale. Our second example is of a gambler following a "double or quits" betting strategy (the word "martingale" in fact comes from an 18th-century term for this strategy). At time $t = 1$, the gambler bets £1 on the toss of a fair coin. If she loses on round $t$, she places a new bet on round $t + 1$ for double the previous stake, but once she has won once, she never bets again. The profit earned on round $t + 1$ is statistically dependent on the history $\mathcal{H}_t$: if the gambler has already won once before this round, she does not bet and $X_{t+1}$ will always be 0; but if she has lost all of the previous $t$ rounds, the stake will be $2^t$, and $X_{t+1}$ will be $2^t$ or $-2^t$ with equal probability. In either case, the conditional expectation of the profit at this timestep is $\mathbb{E}[X_{t+1}|\mathcal{H}_t] = 0$, so the cumulative profit $S_t = \sum_{u=1}^{t} X_u$ is a martingale.

We can construct a martingale out of experimental time series data, if one of the variables in question was randomized. Consider an experiment where a randomized stimulus $R_t$ is delivered on trial $t$, following a probability distribution that may depend on the stimuli and behavior on this and previous timesteps. Suppose we want to test the null hypothesis that brain activity $B_t$ is independent of $R_t$ conditional on history. If we define $X_t = B_t(R_t - \bar{R}_t)$, where $\bar{R}_t = \mathbb{E}[R_t|\mathcal{H}_{t-1}]$, then under the null hypothesis $\mathbb{E}[X_t|\mathcal{H}_{t-1}] = 0$; indeed, $\mathbb{E}[X_t|\mathcal{H}_{t-1}] = \mathbb{E}[B_t(R_t - \bar{R}_t)|\mathcal{H}_{t-1}] = \mathbb{E}[B_t|\mathcal{H}_{t-1}]\mathbb{E}[(R_t - \bar{R}_t)|\mathcal{H}_{t-1}] = 0$. Thus, under the null hypothesis, the timeseries $S_t = \sum_{u=1}^{t} B_u(R_u - \bar{R}_u)$, which measures how brain activity correlates with the deviation of the stimulus relative to its expectation on each round, is a martingale.

## Martingale central limit theorem

The usual central limit theorem says that sums of independent random variables converge to a Gaussian distribution. A version of the central limit theorem also holds for martingales, even though the increments $X_t$ are not statistically independent. The martingale central limit theorem has various forms, but for our purposes the following version suffices:

**Theorem.** *Let $X_t$ be a martingale difference sequence that is uniformly bounded: for some A, $|X_t| \leq A$ for all $t$. Let $\sigma_t^2 = \mathbb{E}[X_t^2|\mathcal{H}_{t-1}]$ be the variance of $X_t$ conditional on history $\mathcal{H}_{t-1}$, and assume $\sum_{t=1}^{\infty} \sigma_t^2 = \infty$ with probability 1. Fix $V > 0$, and let $T$ be the first time when $\sum_{t=1}^{T} \sigma_t^2 \geq V$. Then as $V \to \infty$, the random variable $Z = S_T/\sqrt{\sum_{t=1}^{T} \sigma_t^2}$ converges in distribution to a standard Gaussian.*

**Proof.** Billingsley[3], theorem 35.11, together with the fact that $\sqrt{\sum_{t=1}^{T} \sigma_t^2}/V \to 1$ with probability 1 as $V \to \infty$. □

Roughly speaking, this theorem says that the value of a martingale $S_t$ follows a Gaussian distribution after enough uncorrelated increments have been added to it. Thus, if we can generate a timeseries that under the null hypothesis is a martingale, we can test our null hypothesis by comparing the observed value of $Z = S_T/\sqrt{\sum_{t=1}^{T} \sigma_t^2}$ to critical values of a standard Gaussian. We must however remember two important subtleties.

First, note that the conditions $|X_t| \leq A$ and $\sum_{t=1}^{\infty} \sigma_t^2 = \infty$ are essential. For example, they do not hold for the "double or quits" martingale we described earlier: if the gambler has lost the first $t$ rounds, the stakes of round $t + 1$ is $2^t$, so the $X_t$ are unbounded; also, the cumulative variance is finite with probability 1 since the gambler always stops after her first win. And indeed, this martingale does not converge to a Gaussian distribution: the cumulative profit on time $t$ is 1 with probability $1 - 2^{-t}$, and $1 - 2^t$ with probability $2^{-t}$.

The second subtlety is that the theorem does not say that $S_T/\sqrt{\sum_{t=1}^{T} \sigma_t^2}$ converges to a standard Gaussian as $T \to \infty$. Instead, we need to fix a variance threshold $V$, find the time $T$ when the cumulative variance exceeds $V$, and then the test statistic $Z = S_T/\sqrt{\sum_{t=1}^{T} \sigma_t^2}$ converges to a Gaussian as $V \to \infty$. To see why this makes a difference, suppose that $X_t$ is normally distributed with mean 0 and variance 1 if $S_{t-1} > 0$, and variance $10^{-12}$ if $S_{t-1} \leq 0$. While $S_t$ is positive, it moves around at random with steps of average size 1, so will eventually become negative. However once $S_t$ becomes negative it slows to a virtual standstill. Thus if we pick an arbitrary large time $T$, then $S_T$ is very likely to be negative. However, if we pick the time $T$ when the cumulative variance first crosses a predefined threshold $V$, then $S_T$ is equally likely to be positive as negative.

## Martingale Z-test

We can now describe the martingale Z-test:

1. Define a bounded timeseries $X_t$ whose cumulative sum $S_t$ is a martingale under the null hypothesis. To test if a measured variable $B_t$ is independent of a randomized variable $R_t$ conditional on history, we define $X_t = B_t(R_t - \bar{R}_t)$, where $\bar{R}_t = \mathbb{E}[R_t|\mathcal{H}_{t-1}]$. Note that the distribution of $R_t$ can depend on history, as long as its expectation is known (which it must be if $R_t$ is generated randomly).
2. Fix a cumulative conditional variance threshold $V$. $V$ should be large enough that by the time it is reached, around 30 contributions of approximately equal size have been added together (the same rule of thumb as for using a t-test on non-Gaussian data), but small enough that it is likely to be exceeded before the experiment ends.
3. Compute the cumulative conditional variance $V_t = \sum_{u=1}^{t} \mathbb{E}[X_u^2|\mathcal{H}_{u-1}]$; for correlating $B_t$ with $R_t$ this equals $\sum_{u=1}^{t} B_u^2 \text{Var}[R_u|\mathcal{H}_{u-1}]$. Find the time $T$ where this first exceeds the threshold $V$. If the threshold has not been reached by the end of the experiment, do not reject the null.
4. Define the test statistic $Z = S_T/\sqrt{V_T}$.
5. Reject the null hypothesis if $Z$ lies outside appropriate one- or two-sided critical values of a standard Gaussian distribution.

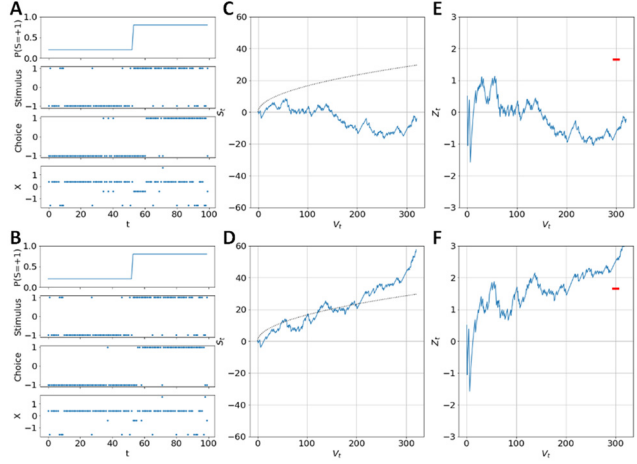

**Figure 1 | Simulations.** Data was simulated for a behavioral task where the probability of a stimulus to appear on the right (S=+1) changes in blocks. **A:** First 100 trials history of block, stimulus, and choice for a simulated subject that infers the correct answer from the history of rewards but not the current stimulus. Bottom row shows martingale difference sequence $X_t$. **B:** Similar plot for a simulated subject whose choices do depend on the current stimulus. **C, D:** Cumulative sum of test statistic $S_t$ vs. cumulative variance $V_t$ for all 500 trials, for the same two subjects. Dashed black line shows equivalent 95% significance level for every possible stopping variance $V_t$. **E, F:** Z-statistic $S_t/\sqrt{V_t}$. Red line shows one-sided critical level at p=0.05 for predefined variance threshold $V = 300$.

## Example

We illustrate the martingale Z-test with the same behavioral task described on page 1. We simulate two subjects' behavior for 500 trials. The first subject (Figure 1A) cannot directly see the stimulus but is able to infer which side previous stimuli appeared on from delivery of rewards, and so learns which side is likely to be rewarded in the current block (see Methods for details). The second subject's choices are also directly influenced by the stimulus on trial $t$; observe that this subject switches their choices sooner after the block changes (Figure 1B). Naïve application of Fisher's exact test for association of stimulus and choice gives significance at $p = 10^{-9}$ and $10^{-28}$ for the two subjects, even though the first subject cannot see the stimulus.

To apply the martingale Z-test, we construct the martingale difference sequence $X_t = C_t(A_t - \bar{A}_t)$, where $C_t = \pm 1$ is the subject's choice on trial $t$, $A_t = \pm 1$ is the stimulus shown on this trial. The history-conditional expectation $\bar{A}_t$ is easily shown to be $\pm 0.6$ depending on the block, and so $X_t = \pm 0.4$ when the stimulus most likely for the current block appears, with positive sign for a correct choice and $X_t = \pm 1.6$ when the stimulus less likely for this block appears, again with positive sign for correct. The conditional variance $\sigma_t^2 = \mathbb{E}[X_t^2|\mathcal{H}_{t-1}]$ is easily shown to be 0.64 on all trials, and we choose a predefined cumulative conditional variance threshold of $V = 300$. Plotting the martingale sum $S_t$ vs the cumulative variance $V_t$, we see that for the simulated subject that cannot see the stimulus, $S$ follows a random walk without drift (Figure 1C), whereas there is positive drift for the subject that can see the stimulus (Figure 1D). Applying the one-sided critical value of 1.65 at the time when cumulative variance first exceeds $V = 300$, we reject the null only for the subject that can see the stimulus (Figure 1E, F).

## Choosing V

An unusual feature of this test is that one must pre-select a cumulative variance threshold $V$ before analyzing the data. The larger $V$ is, the more power the test has, but $V$ should not be so large that it risks not being reached by the end of the experiment. As always when statistical tests have free parameters, one must avoid "p-hacking" by selecting a specific $V$ post-hoc that gives significance after examining plots like Figure 1E, F.

One might be tempted to simply compute the Z statistic on the final timepoint of the experiment. However, this strategy can also lead to false rejection of a valid null. For example, if $X_t$ is a Gaussian distribution whose variance decreases to zero when $Z_{t-1}$ exceeds the test's critical value, then $S_t = \sum_{u=1}^{t} X_u$ is a martingale, but $S_t$ and $Z_t$ will both freeze as soon as the test becomes significant, and the overall test will show significance if $Z_t$ is significant at any time point; one might say the test p-hacks itself. Defining $V$ ahead of time avoids this confound: if the step size drops to zero before cumulative variance reaches $V$, the null not rejected.

Other test statistics, based on first-crossing times of Brownian motion[4,5] (which the function $S(V)$ will approximate), might be able to provide an alternative to fixing $V$, but we have not yet investigated this possibility.

## Tangent sequence randomization

The martingale Z-test is approximately equivalent to a randomization test. Randomization tests often require fewer assumptions than parametric tests, so one might hope that this randomization test avoids requirements such as approximately equal bounded increments and a predefined variance threshold. However this is not the case, and there is no advantage to this randomization method over the actual Z-test. Nevertheless, it is still illustrative to see why it works.

In a standard randomization test, one compares a test statistic computed from the observed data to an ensemble of test statistics computed from synthetic data that under the null hypothesis follow the same probability distribution as the actual data. This produces a conservative test: if the null holds, false positive results cannot occur with probability exceeding the stated p-value. In the present context, we cannot resample from the actual distribution, which is unknown. Nevertheless, we can still come up with a null ensemble that has approximately the same distribution as the full data, under the same assumptions as the martingale Z-test.

In this approach, we obtain a null distribution by repeatedly replacing the observed sequence $X_t$ with a tangent sequence $X_t'$. In a tangent sequence, $X_t'$ is randomly drawn for each time $t$ from the distribution $\mathbb{P}[X_t|\mathcal{H}_{t-1}]$. The history $\mathcal{H}_{t-1}$ means the actual observations until time $t-1$, not the random samples. Tangent sequences have a different character to actually observed sequences. For example, in the double-or-quits martingale, any history contains exactly one win. But a tangent sequence to an observed sequence that lasted $T$ trials will contain $n$ wins with probability $2^{-n}\binom{T}{n}$.

To apply the tangent sequence randomization test, we compare $S_T = \sum_{t=1}^T X_t$ to a null ensemble obtained by repeatedly constructing tangent sequences $X_t'$, and summing them to produce $S_T' = \sum_{t=1}^T X_t'$. Conditional on the observed data, $X_t'$ is a sequence of independent random variables with mean 0 and variances $\sigma_t^2$. Under conditions where the martingale central limit theorem applies to $X_t$, it also applies to $X_t'$, so $S_T' \sim N(0, V_T)$. Tangent randomization thus rejects the null if $S_T$ exceeds a critical point on this Gaussian distribution, and thus is equivalent to the martingale Z test under the conditions where the martingale Z test holds.

If the conditions for the martingale Z test do not hold, the tangent sequence randomization test may not be valid. For example, consider a gambler playing a double-or-quits strategy, but where the probability of a win on each trial is $\alpha$ and the odds are fair: $\alpha^{-1} - 1$ for a win, $-1$ for a loss. The gambler's total cumulative profit is a martingale, which takes the final value $S_T = 2^{T-1}\alpha^{-1} - 2^T + 1$ for a win on trial $T$. In a tangent sequence, the total profit is $S_T' = \sum_{t=1}^T 2^{t-1} X_t'$, where $X_t'$ is a sequence of independent variables equal to $\alpha^{-1} - 1$ with probability $\alpha$ and $-1$ with probability $1 - \alpha$. Now if $\alpha \leq \frac{1}{2}$ then $S_T' < 1$ if and only if $X_t' = -1$, which occurs with probability $1 - \alpha$. Thus, the observed test statistic exceeds the $(1 - \alpha)^{th}$ quantile of the null distribution with probability 1, and we always reject the null hypothesis with $p = \alpha$ even though the null is valid.

# Summary

We have described a statistical test for independence of two time series conditional on history, where the history-conditional distribution $\mathbb{P}[R_t|\mathcal{H}_{t-1}]$ of one of the series known. It works by defining a test statistic $S_t = \sum_{u=1}^t X_t$, where $X_t = B_t(R_t - \mathbb{E}[R_t|\mathcal{H}_{t-1}])$. $S_t$ is a martingale if $B_t$ and $R_t$ are independent conditional on $\mathcal{H}_{t-1}$. If the $X_t$ are bounded and the cumulative conditional variance $V_t = \sum_{u=1}^t B_t^2 Var[R_t|\mathcal{H}_{t-1}]$ is sufficiently large, then $Z_t = S_t/\sqrt{V_t}$ is close to a standard Gaussian, and the null can be tested by a Z-test on $Z_T$, where $T$ is the first time that $V_t$ exceeds a predefined threshold.

# Methods

In the simulations of Figure 1, an unobserved "block" variable $B_t = \pm 1$ switched sign every 50-100 trials (uniform distribution). The stimulus $A_t = B_t$ with probability 0.8 and $A_t = -B_t$ with probability 0.2. To simulate the subject's choices, "reward learning" and "habit learning" variables[6] $\rho_t$ and $h_t$ were integrated according to the history of choices $C_t$ and rewards $R_t$:

$$\rho_{t+1} = \beta\rho_t + C_t R_t$$
$$h_{t+1} = \beta h_t + C_t$$

where the decay constant $\beta = 0.65$. The log-odds of the subjects choice is a linear function of these two variables and of the stimulus:

$$\log\frac{\mathbb{P}(C_t = +1)}{\mathbb{P}(C_t = -1)} = W_\rho \rho_t + W_h h_t + W_A A_t$$

The parameters $W_\rho$ and $W_h$ both took the value 1, while $W_A$ was zero for the "stimulus-blind" subject (Figures 1A,C,E) and 1 for the "stimulus-visible" subject (Figures 1B,D,F).

Code for the simulations in Figure 1 is available at https://colab.research.google.com/drive/1sfdZJGPqNm1trD5QVpRZtAldsaJCsfva.

1. Harris, K. D. Nonsense correlations in neuroscience. *bioRxiv* 2020.11.29.402719 (2020) doi:10.1101/2020.11.29.402719.
2. The International Brain Laboratory *et al*. Standardized and reproducible measurement of decision-making in mice. *eLife* **10**, e63711 (2021).
3. Billingsley, P. *Probability and Measure*. (Wiley, 1995).
4. Nádas, A. Best Tests for Zero Drift Based on First Passage Times in Brownian Motion. *Technometrics* **15**, 125–132 (1973).
5. Sato, S. Evaluation of the First-Passage Time Probability to a Square Root Boundary for the Wiener Process. *Journal of Applied Probability* **14**, 850–856 (1977).
6. Miller, K. J., Shenhav, A. & Ludvig, E. A. Habits without values. *Psychol Rev* **126**, 292–311 (2019).